\documentstyle[aps,prl,twocolumn,epsf]{revtex}

\begin{document}

\hyphenation{Sr}

\twocolumn[
\hsize\textwidth\columnwidth\hsize\csname@twocolumnfalse\endcsname
\draft

\title{Negative Magnetoresistance of Bi$_2$Sr$_2$CuO$_x$ Single Crystals in a
Strong Magnetic Fields.}
\author{S.I.Vedeneev,$^{1,2}$ A.G.M.Jansen,$^{1}$, B.A.Volkov,$^{2}$, and P.Wyder$^{1}$}
\address{$^1$High Magnetic Field Laboratory,
Max-Planck-Institut f\"{u}r Festk\"{o}rperforschung / Centre
National de la Recherche Scientifique,
B.P. 166, F-38042 Grenoble Cedex 9, France}
\address{$^{2}$P.N. Lebedev Physical Institute, Russian Academy
of Sciences, SU-117924 Moscow, Russia}


\maketitle

\begin{abstract}
Magnetoresistance (MR) in the out-of-plane resistivity $\rho _c$ for the
normal state of the one-layer high-quality Bi$_2$Sr$_2$CuO$_x$
single crystals under various dc magnetic fields up to $28$ T over the
temperature region $6-100$ K has been measured. We observed the anomalously
large negative longitudinal MR up to $60\%$. At low temperatures the
normal-state MR in contrast to the MR in mixed state is independent of the
direction of the current relatively to the field direction suggesting
uniquely the spin dominated origin of that. The magnitude of the MR is
activated in magnetic field and temperature. We interpret the activated form
of $\rho _c$ and the negative MR in terms of 2D stacked alternating metallic
and dielectric layers assuming the tunneling between CuO$_2$ planes. If the
main fluctuations inside CuO$_2$ planes have magnetic origin, the magnetic
field suppresses these fluctuations leading to the uniform spin orientation.
In this case the interlayer current will be enhanced well.
\end{abstract}

]

1. The magnetotransport properties of the layered high-$T_c$ superconductors
(HTSC) are characterised by anomalous quasi-two-dimensional (2D) states,
which are studied very extensively in recent years$^1$. One of unusual
features of the HTSC normal state properties is the coexistence of the
metallic in-plane resistivity $\rho _{ab}$ and the ''semiconducting''
out-of-plane resistivity $\rho _c$ (e.g. Ref.$^{2,3}$). Recently, such
behaviour of the resistivities $\rho _{ab}$ and $\rho _c$ were measured by
Ando et al.$^4$ in La-doped Bi$_2$Sr$_2$CuO$_y$ ($T_c=13$ K) down to
temperature as low as $T/T_c\sim 0.04$. The latter implies a 2D confinement
and is incompatible with a Fermi-liquid behaviour. The small-value ($\sim
1\% $) negative out-of-plane magnetoresistance (MR) around $T_c$ for Bi$_2$Sr%
$_2$CaCu$_2$O$_8$ and YBa$_2$Cu$_3$O$_7$ and was explained by a fluctuation
conductivity or in terms of a pseudogap in spin system that was reduced by
magnetic field$^{3,5-7}$. A small negative isotropic MR was observed also in
underdoped La$_{2-x}$Sr$_x$CuO$_4$$^8$.

The normal-state properties essentially depend on a carrier concentration or
doping also. In recent experiments the unusual field dependence of the large
positive $c$-axis MR of the underdoped cuprate YBa$_2$Cu$_4$O$_8$ has been
assigned to the magnetic field decouples of the chains resulting in a 3D to
2D crossover in the transport behaviour of the whole system$^9$. Yoshizaki
et al.$^{10}$ have observed positive and negative out-of-plane MR up to $2\%$
over the temperature region $35-200$ K for magnetic fields $0-17$ T in
underdoped Bi$_2$(SrLa)$_2$CuO$_{6+\delta }$. Although magnitude of the
negative longitudinal MR was about five times larger than that for the
transverse MR they have proposed that data indicated the Lorentz-force
independent spin-dominated origin of the MR. These results were discussed
from a view point of pseudo-spin-gap formation. Very recently a change of
sign from positive near $T_c$ to negative above $T_c$ in the out-of-plane
magnetoconductivity of the Tl$_2$Ba$_2$Ca$_2$Cu$_3$O$_{10+\delta }$ compound$%
^{11}$ was again explained by the density-of-states fluctuations. In the
same time the one-layer (Hg,Cu)Ba$_2$CuO$_{4+\delta }$ did not show any
anomaly in its magnetoconductivity$^{11}$. At present no single theory has
been able to account for all the anomalies found in the normal-state
properties of HTSC.

In this paper, we describe the experiments to study the $c$-axis resistance
in normal state of high-quality nondoped Bi$_2$Sr$_2$CuO$_{6+\delta }$
(Bi2201) single crystals under various dc magnetic fields up to $28$ T over
the temperature region $6-100$ K. At low temperatures the almost isotropic
negative out-of-plane MR up to $60\%$ was observed.

\smallskip

2. The Bi2201 single crystals were grown by KCl-solution-melt method$^{12}$.
A temperature gradient along a crucible results in formation a big closed
cavity inside the solution-melt. Number of the crystals which share the
common properties reached several tens in the cavity. The quality of the
crystals was verified by the measurements of the $dc$ resistance, $ac$
susceptibility, X-ray diffraction and energy dispersive X-ray microprobe
analysis. Our crystals showed the X-ray rocking curves width about $%
0.1^{\circ }-0.3^{\circ }$. The two crystals with $T_c=9.5$ K (midpoint) and
$\Delta T_c\simeq 1~$ K were investigated. Dimensions of the crystals were $%
0.5~mm\times 1~mm\times 3~\mu $m (\#1) and $0.5~mm\times 1~mm\times 10~\mu m$
(\#2). A four-probe contact configuration with symmetrical position of the
low-resistance contacts ($<1\Omega $) on both $ab$-surfaces of the sample
was used for the measurements of the in-plane and out-of-plane resistances.
The measured resistances was transformed to $\rho _c$ and $\rho _{ab}$ using
the crystal dimensions. In zero magnetic field the samples showed nearly
linear temperature dependence $\rho _{ab}(T)$ which saturates below $20$~K
to a residual resistivity $50$ and $80~\mu \Omega \cdot $cm. The
out-of-plane resistivity of the single crystals $\rho _c(T)$ over the
temperature region $T=10-300$~K showed the conventional ''semiconducting''
normal-state behaviour of $\rho _c$ in layered Bi-family and was in
reasonable good agreement with a power law $T^{-\alpha }$ with $\alpha =1.65$
(crystal \#1) and $1.3$ (crystal \#2) without a linear-$T$ term. The $\rho
_c $ values at $T=100$ K of the thin and thick samples are equal $2.7$ and $%
13.5 $ m$\Omega \cdot $cm, respectively. The anisotropy ratio $\rho _c/\rho
_{ab}$ was nearly $5\times 10^4$ at low temperatures. By measuring the Hall
coefficient $R_H$ in the crystals we determined the carrier density $%
n=4.8\times 10^{21}$cm$^{-3}$. The crystals were studied with the magnetic
field ${\bf H}$ in longitudinal (${\bf H\parallel c\parallel J}$) and
transverse (${\bf H\parallel ab\perp J}$) configurations.

\smallskip

3. Fig.1 displays the field dependence of the longitudinal $\rho _c$ (a) and
MR, $\Delta \rho _c/\rho _{c0}$, (b) as $H$ is parallel to $c$-axis for the
thin sample \#1 at different temperatures just below $T_c$ and above $T_c.$
(At $T<T_c$ the peak value of $\rho _c(H)$ was selected as $\rho _{c0}$).
The out-of-plane resistivity is decreasing with increasing magnetic field,
which is consistent with the preceding papers.
\vspace{-0.7cm}
\begin{figure}[htbp]
\begin{center}
 \epsfxsize=75mm
 $$\epsffile{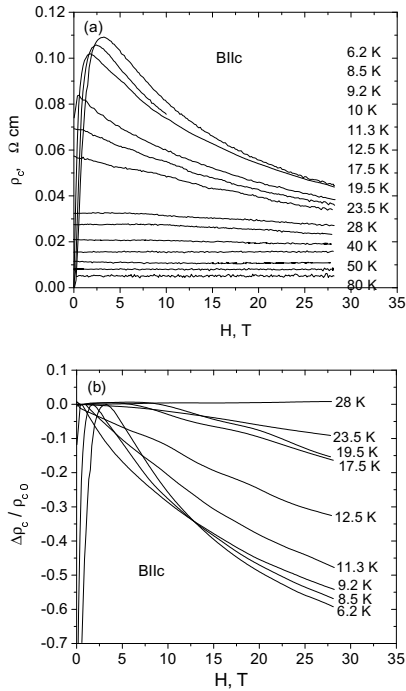}$$
\end{center}
\caption{The field dependence of the out-of-plane resistivity  $\rho _c$ (a)
and longitudinal MR, $\Delta \rho _c/\rho _{c0}$, (b) as $H$ is parallel to $%
c$-axis for the sample \#1 at different temperatures just below $T_c$ and
above $T_c.$ (At $T<T_c$ the peak value of $\rho _c(H)$ was selected as $%
\rho _{c0}$).}
\label{fig1}
\end{figure}
However there are two
distinguishing features to note from this figure. First, at low temperatures
the $\rho _c$ decreases too rapidly as compared with other works. For
example, at $T\approx 6$ K the $\rho _c$ decreases its value more than by a
factor of $2.5$. With increasing temperature the MR decreases and becomes
positive above $\approx 26$ K. In the very high fields $\rho _c$ shows
tendency to the saturation. We have observed the clearly defined saturation
of the out-of-plane MR after its twofold decreasing in our crystals at $%
T=1.9-4.2$ K in the pulsed magnetic fields the longitudinal geometry at $%
40-50$~T$^{13}$. However, the precise measurement of $\rho _c$ in the pulsed
magnet was found a difficult task due to very low out-of-plane resistance of
our crystals and higher noise level. It should be noted that Ando et al.$^4$
have observed the $10\%$ negative out-of-plane MR in La-doped Bi$_2$Sr$_2$CuO%
$_{y\text{ }}$ at $T=0.8$ K by suppressing superconductivity with pulsed
magnetic fields. They have found that the magnetic-field dependence of the
negative MR is approximately linear and there is no sign of saturation up to
$60$ T. The second peculiarity is the extremely large negative MR which was
observed in the parallel configuration where the macroscopic Lorentz force
should be absent.

\vspace{-0.7cm}
\begin{figure}[htbp]
\begin{center}
 \epsfxsize=75mm
 $$\epsffile{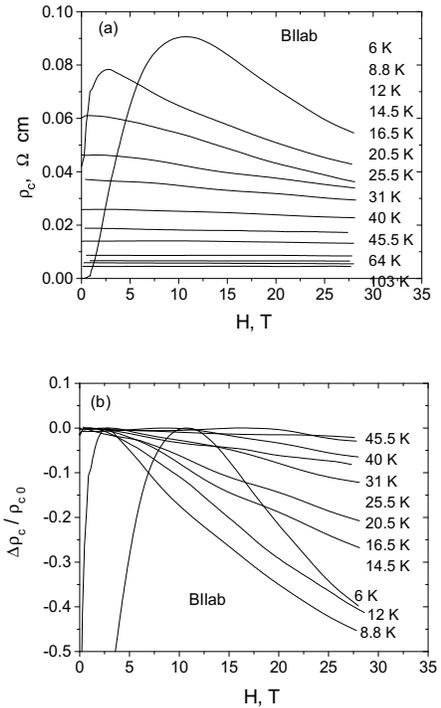}$$
\end{center}
\caption{The out-of-plane resistivity $\rho _c$ (a) and $\Delta \rho _c/\rho
_{c0}$ (b) in the transverse configuration (${\bf H\parallel ab}$, ${\bf %
J\parallel c}$) as a function of $H$ for the same sample \#1 at selected
temperatures.}
\label{fig2}
\end{figure}

The $\rho _c$ (a) and $\Delta \rho _c/\rho _{c0}$ (b) in the transverse
configuration (${\bf H\parallel ab}$, ${\bf J\parallel c}$) are plotted in
Fig.2 as a function of $H$ for the same sample \#1 at selected temperatures.
It is apparent that a very large negative out-of-plane MR is present in the
normal-state once again. The difference in the normal-state magnitudes of $%
\Delta \rho _c/\rho _{c0}$ in Fig.2 and Fig.1 is no more than several
percent. One can also see that the normal-state MR in contrast to the MR in
mixed state is almost independent of the magnetic field direction with
respect to the current direction. The larger resistive onset field at $T<T_c$
is a direct consequence of the large anisotropy of upper critical field in
Bi2201. The critical field in this direction is much larger than that when
the field is perpendicular to the $ab$-plane. While
Lorentz-force-independent MR in normal state uniquely indicates the spin
dominated origin of that.

Similar behaviour of the $\rho _c$ and $\Delta \rho _c/\rho _{c0}$ from
second crystal \#2 was obtained exept that at high magnetic fields the
magnitude of $\Delta \rho _c/\rho _{c0}$ was $\approx 2$ times smaller
previous one. The crystals \#1 and \#2 have the same nominal composition,
the same $T_c,$ but were grown in different crucibles. The fact that $\rho
_c $ and MR of the crystals significantly differ presumably reflects varying
disorder along $c$-axis, which is believed to be due to additional
insulating layers in the thick sample. Our studies shown that those layers
existed generally in the crystals thicker than $3-5$ $\mu $m.

\vspace{-0.7cm}
\begin{figure}[htbp]
\begin{center}
 \epsfxsize=50mm
 $$\epsffile{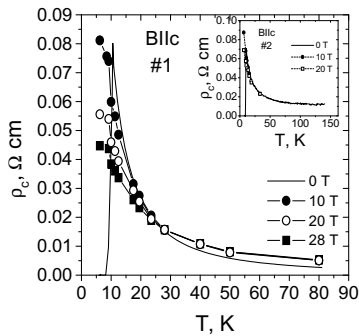}$$
\end{center}
\caption{The temperature dependencies of the out-of-plane resistivity $\rho
_c(T)$ in sample \#1 over the temperature region $6-80$ K at zero-field
curve (solid line) together with values of $\rho _c(T)$ at $10,$ $20$, and $%
28$ T extracted from Fig.1(a). The inset shows similar $\rho _c(T)$ at
zero-field (solid line), at $10$, and $20$ T data for sample \#2.}
\label{fig3}
\end{figure}

The temperature dependencies of the out-of-plane resistivity in sample \#1
over the temperature region $6-80$ K is represented in Fig.3, where we show
the zero-field $\rho _c(T)$ curve (solid line) together with values of $\rho
_c(T)$ at $10,$ $20$, and $28$ T extracted from Fig.1(a). The $\rho _c(T,H)$
curves intersect the zero-field $\rho _c(T)$ curve at $T\approx 26$ K where
the MR shows a change of sign as the temperature is increased. The inset
shows similar $\rho _c(T)$ at zero-field (solid line), at $10$, and $20$ T
data for sample \#2.

In the quasi-classical model the longitudinal MR arises from a
spin-dependent scattering which is independent of the field direction. This
MR in conventional metals is positive, very small ($\sim 10^{-3}$) and
increases quadraticaly with $H$. In cited above works$^{7,10\text{ }}$the
negative out-of-plane MR is quadratic of the magnetic field in the wide
range of the field up to $14$ T to a good approximation or that is linear up
to $60$ T$^4$. Our $\rho _c$ data can be well fitted to the functional form $%
\rho _c(H)=A_0+A_1\exp (-H/bT)$, where $H$ and $T$ are in T. The constant $b$
equals $\approx 2.8,$ with the Zeeman field taken as $H_Z=k_BT/g\mu _B$ $%
\approx 0.74~$T at $1~$K ($g=2$ and $\mu _B$ are $g$ factor and Bohr
magneton). The anomalously large negative longitudinal MR has been observed
more early in the transition metal dichalcogenides$^{14}$ which have a layer
type structure. Fukuyama and Yosida$^{15,16}$ have explained this phenomenon
on basis of a variable-range hopping mechanism in Anderson localized states$%
^{17}$. The application of the magnetic field introduces Zeeman shifts of
each eigenstate dependent on the spin directions and causes the repopulation
among the localized states. The energy difference decrease with increasing
field and the interplane conduction increases exponentially$^{16}$. If we
assume that this scenario is correct for Bi2201 compound also then it may
explain the negative longitudinal MR. However, the magnitude of the observed
MR is too large to be considered for this model and the field value at which
the MR should fall off sharply is more less than we observe.

We have attempted to describe qualitatively our data in terms of 2D stacked
alternating metallic and dielectric layers assuming the tunneling of
electrons between the adjacent CuO$_2$ planes. Based on the resistance
anisotropy ($\rho _c/\rho _{ab}\sim 10^4-10^5$) it is believed that a
transition amplitude of the charge carriers between the planes is two order
of magnitude less than that within the CuO$_2$ layer over the same distance.
As an added argument of the 2D charge transport in the Bi2201 crystals is a
slight negative in-plane MR $\Delta \rho _{ab}/\rho _{ab0}$, which we
observed only when the field is applied perpendicular to the CuO$_2$ layers.
In this case the magnetic field dependence of in-plane MR is partially
described by weak localization theory. Observed here the extremely large
negative out-of-plane MR is independent of the magnetic field direction,
suggesting the spin-dominated origin of MR.

We will proceed from the assumptions that: (i) - the current $J$ along $c$%
-axis determines by the electron tunneling between the neighbouring metallic
weakly linked layers with the lower-order transition amplitude equals T ($%
J\sim $ $\mid $T$^2\mid $); (ii) - the crystal is perfected, then, the
charge carriers transport between the layers should be elastic with
conserving of the quasiparticle momentum along the layers and spin; (iii) -
the planar density of states in the vicinity of a Fermi level is free from
the singularities. From the last assumption it follows that in the absence
of an interaction between electrons in each layer the $c$-axis current has
not to depend on the magnetic field. Really, though a spin-splitting of the
2D-bands leads to an electron density redistribution among the bands, the
total current is given by the sum of both spin components remains constant.
It follows that the negative MR can be associated with the electron gas
unideality only which should depend on the magnetic field. In terms of an
one-particle Green functions it is inferred that a renormalization constant $%
Z$ in the definition of the electron Green function in the layer $G=Z$ $/$ $%
(\omega -\varepsilon _k-U)$ is less than $1$, where $\varepsilon _k$ is a 2D
electron spectr in the layer and $U$ is a layer potential. (For the ideal
gas $Z=1$).

According to our assumptions (i) and (ii), the tunnel current is determined
by a convolution of the Green function in the adjacent CuO$_2$ layers.
Vertexes of the convolution are transition amplitudes among the layers which
equal T. Hence one can expect that the interlayer current $J$ is
proportional to $Z^2$. If the magnitude of $Z$ increases with increasing
field, the current $J$ will be enhanced well. This is possible when main
fluctuations determining $Z$ value in the layer have a magnetic origin. In
this case the magnetic field suppresses these fluctuations leading to the
uniform spin orientation. It is evident that in an intense fields saturation
of the magnetic field dependence of the tunnel conductance is expected when
all spins are oriented in the same direction and the temperature dependence
of MR has activated character.

Quantitative calculation of $Z(H,T)$ behaviour essentially depends on
choosing of a model of the  magnetic correlations in CuO$_2$ layer. These
correlations may be caused by a band magnetism, interaction of the copper
magnetic moments with the band electrons, various paramagnons and so on. The
different theoretical models should be compared with the MR experimental
data in the further in order to make clear a nature and role of the magnetic
interactions in the superconductivity of high-$T_c$ cuprates.

To summarize, we measured the MR in the out-of-plane resistivity for the
normal state of the one-layer high-quality Bi$_2$Sr$_2$CuO$_{6+\delta }$
single crystals under various dc magnetic fields up to $28$ T over the
temperature region $6-100$ K. We observed the anomalously large negative
longitudinal MR up to $60\%$. At low temperatures the normal-state MR in
contrast to the MR in mixed state is independent of the direction of the
current relatively to the field direction suggesting uniquely the spin
dominated origin of that. The magnitude of the MR is activated in magnetic
field and temperature. With increasing temperature the MR decreases and
becomes positive above $\approx 26$ K. In the very high fields MR shows
tendency to the saturation. We interpret the activated form of $\rho _c$ and
the negative MR in terms of 2D stacked alternating metallic and dielectric
layers assuming the tunneling between CuO$_2$ planes. If the main
fluctuations inside CuO$_2$ planes have magnetic origin, the magnetic field
suppresses these fluctuations leading to the uniform spin orientation. In
this case the interlayer current  will be enhanced well.

\smallskip

Two of us (S.I.V. and B.A.V.) were partially supported by {Russian Ministry
of Science and Technical Policy in the frame of the Program Actual Problems
of Condensed Matter Physics Projects N96001, N96081 and by the Russian
Foundation for Basic Research Projects N99-02-17877, N99-02-17360}

\smallskip
{\bf Reference}

\noindent 1. Wu Liu, T.W.Clinton, A.W.Smith, C.J.Lobb, Phys. Rev. B {\bf 55}%
, 11802 (1997), and references cited therein.\newline
2. S.Martin, A.T.Fiory, R.M.Fleming, L.F.Schneemeyer, J.V.Waszczak, Phys.
Rev. ${\bf B41}$, 846 (1990).\newline
3. G.Briceno, M.F.Crommie, A.Zettl, Phys. Rev. Lett. ${\bf 66}$, 2164 (1991).%
\newline
4. Y.Ando, G.S.Boebinger, A.Passner, N.L.Wang, C.Geibel, F.Steglich, Phys.
Rev. Lett. ${\bf 77}$, 2065 (1996).\newline
5. S.I.Vedeneev, A.G.M.Jansen, P.Samuely, V.A.Stepanov, A.A.Tsvetkov,
P.Wyder, Phys. Rev., ${\bf B49}$, 9823 (1994).\newline
6. K.Nakao, K.Takamuku, K.Hashimoto, N.Koshizuka, S.Tanaka, Physica ${\bf %
B201}$, 262 (1994).\newline
7. Y.F.Yan, P.Matl, J.M.Harris, N.P.Ong, Phys. Rev., ${\bf B52}$, R571
(1995).\newline
8. Y.Ando, G.S.Boebinger, A.Passner, T.Kimura, K.Kishio, Phys. Rev. Lett. $%
{\bf 75}$, 4662 (1995).\newline
9. N.E.Hussey, M.Kibune, H.Nakagawa, N.Miura, Y.Iye, H.Takagi, S.Adachi,
K.Tanabe, \text{Phys. Rev. Lett. {\bf 80}, 2909 (1998). }\newline
10. R.Yoshizaki, H.Ikeda, Physica ${\bf C271}$, 171 (1996). \newline
11. A.Wahl, D.Thopart, G.Villard, A.Maignan, V.Hardy, J.C.Soret, L.Ammor,
A.Ruyter, Phys. Rev., ${\bf B59}$, 7216 (1999).\newline
12. J.I.Gorina, G.A.Kaljushnaia, V.P.Martovitsky, V.V.Rodin, N.N.Sentjurina,
Solid State Commun. ${\bf 108}$, 275 (1998).\newline
13. These measurements were done in Service National Des Champs Magnetique
Pulses Laboratory, Toulouse, France. \newline
14. N.Kobayashi, Y.Muto, Solid State Commum. ${\bf 30}$, 337 (1979).%
\newline
15. H.Fukuyama, K.Yosida, J.Phys.Soc.Jpn. ${\bf 46}$, 102 (1979).\newline
16. H.Fukuyama, K.Yosida, J.Phys.Soc.Jpn. ${\bf 46}$, 1522 (1979).\newline
17. P.W.Anderson, Phys. Rev., ${\bf 109}$, 9823 (1958).

\end{document}